\documentclass[preprint,12pt]{elsarticle}

\usepackage{amssymb}
\usepackage{bm}
\usepackage{amsthm}
\usepackage{dsfont}
\usepackage{algorithm}
\usepackage{algorithmic}
\usepackage{subfigure}

\usepackage{lineno}
\usepackage[dvipdfm,  
            pdfstartview=FitH,
            CJKbookmarks=true,
            bookmarksnumbered=true,
            bookmarksopen=true,
            colorlinks, 
            pdfborder=001,   
            linkcolor=green,
            anchorcolor=green,
            citecolor=green
            ]{hyperref}

\newcommand{\eqn}[2]
{
\begin{equation}#1
#2
\end{equation}
}

\journal{}

\begin{document}

\begin{frontmatter}



\title{Combination of Multiple Bipartite Ranking for Web Content Quality Evaluation}

\author{Xiao-Bo Jin\corref{cor1}}
\ead{xbjin9801@gmail.com}
\address{School of Information Science and Engineering\\Henan University of Technology\\
 Zhengzhou,Henan, 450001, P. R. China}
 \cortext[cor1]{Corresponding author. Tel: +86-0371-67756527}
\author{Guang-Gang Geng}
 \ead{gengguanggang@cnnic.cn}
 \address{Computer Network Information Center\\Chinese Academy of Science\\Beijing,100190,P.R. China}
\author{Dexian Zhang}
\ead{zdxzzit@hotmail.com}
\address{School of Information Science and Engineering\\Henan University of Technology\\
 Zhengzhou,Henan, 450001, P. R. China}


\begin{abstract}
Web content quality estimation is crucial to various web
content processing applications. Our previous work applied Bagging + C4.5 to achive the best results on the ECML/PKDD Discovery Challenge 2010, which is the comibination of many point-wise rankinig models. In this paper, we combine multiple pair-wise bipartite ranking learner to solve the multi-partite ranking problems for the web quality estimation. In encoding stage, we
present the ternary encoding and the binary coding extending each rank value to
$L - 1$ (L is the number of the different ranking value). For the decoding, we
discuss the combination of multiple ranking results from multiple bipartite
ranking models with the predefined weighting and the adaptive weighting. The experiments
on ECML/PKDD 2010 Discovery Challenge datasets show that \textit{binary coding}
+ \textit{predefined weighting} yields the highest performance in all four
combinations and furthermore it is better than the best results reported in
ECML/PKDD 2010 Discovery Challenge competition.
\end{abstract}

\begin{keyword}
Web Content Quality Estimation \sep Multipartite Ranking \sep Bipartite Ranking \sep Encoding Design \sep Decoding Design
\end{keyword}

\end{frontmatter}


\section{Introduction}
In the past, most data quality measures were developed on an ad hoc
basis to solve specific problems, and fundamental principles
necessary for developing stable metrics in practice were
insufficient \cite{Pipino2002}. In the research of Web content
quality assessment, computational models that can automatically
predict the Web content quality should be focused on.

Web spam can significantly deteriorate the quality of search engine
results, but high quality is more than just the opposite of Web
spam. ECML/PKDD 2010 Discovery Challenge (DC2010) aims at more
aspects of the Web sites. DC2010 wants to develop site-level
classification for the genre of the Web sites (editorial, news,
commercial, educational, ``deep Web" or Web spam and more) as well
as their readability, authoritativeness, trustworthiness and
neutrality \cite{Benczur20101}.

The algorithms of learning to rank are traditionally classified as
three categories. In the simplest point-wise approach, the instances
are assigned a ranking score as the absolute quantity using
classical regression or classification techniques
\cite{Crammer2002,Li2007}. In the pairwise approach, the
order of pairs of instances is treated as a binary label and learned
by a classification method (e.g. RankSVM \cite{Joachims2002}). RankBoost
\cite{Freund2003} maintains $n$ weak ranking functions where each
function can orders the instances and then combine the ranking
functions into a single ranking. Finally, the most complex list-wise approaches
\cite{Valizadegan2009} try to directly optimize a ranking-specific
evaluation metric (e.g. Normalized Discounted Cumulative Gain,
NDCG).

In DC2010 2010, Geng et. al\cite{Geng2010} estimate the web quality with the weighted output of bagging of C4.5 to achieve the best results among all submitted reports, which can be regarded as the combination of the point-wise ranking method. However, the pair-wise ranking models often shows the better performance than the point-wise ones and less model complexity than the list-wise ones. In our study, we research on combining multiple binary pair-wise ranking with the efficient ranking coding and decoding for the web quality assessment, which converts the multi-partite
ranking problem into multiple bipartite ranking problems. Specially, we use bipartite RankBoost as the base learner. In encoding stage, we
present the ternary encoding and the binary coding extending each rank value to
the vector with $L - 1$ dimensions ($L$ is the number of the ratings). For the
decoding, we discuss the combination of multiple ranking results from multiple
bipartite ranking models with the predefined weighting and the adaptive
weighting. DC2010 \cite{Benczur20101}
provide a chance to validate our algorithm called MultiRanking.ED, where the tasks is to rank
the webpages from three international languages (English,French,Germany)
according to their quality. The results show that MultiRanking.ED achieves the better results. In particular,
\textit{binary coding} + \textit{predefined weighting} yields the highest
performance in all four combinations and overpasses the best results with
Bagging + C4.5 \cite{Breiman1996,Quinlan1993} in DC2010 competition by our team \cite{Geng2010}.

The reminder of the paper is organized as follows: Section 2 gives the
multi-partite ranking problem and RankBoost algorithm; Section 3 presents
multi-parite ranking with the efficient encoding-decoding; Section 4 provides the experimental results; the last section concludes the paper and the future work.

\section{Multipartite Ranking}
\subsection{Framework of Multipartite Ranking}
In the bipartite ranking problem, given the dataset $S = \{S_{+},S_{-}\}$ from
the instance space $X$ where $S_{+} = \{\mathbf{x}^{+}_i\}_{i = 1}^m$ and
$S_{-} = \{\bm{x}^{-}_j\}_{j = 1}^n$, then the objective of the ranking
algorithm is to minimize the expected empirical error on the function $f: X
\rightarrow \mathbb{R}$:
\begin{equation}\label{eqn:r_f}
R_{I}(f) = \frac{1}{mn}\sum_{i = 1}^m \sum_{j = 1}^n I[f(\bm{x}^{+}_i) <
f(\bm{x}^{-}_j)],
\end{equation}
where $I[\cdot]$ is the indicator function. It is worthy to note that we can
compute the area under the ROC curve (AUC) \cite{Fawcett2006} as $1 -
R_{I}(f)$. As the direct minimization of the 0-1 loss is computationally
intractable, the ranking algorithm minimizes the convex upper bound of the
expected empirical error \footnote{The expected empirical error also includes a
regularization item in some ranking algorithm such as RankSVM.}:
\begin{equation}\label{eqn:bipartite_error}
R_{\phi}(f) = \frac{1}{mn}\sum_{i = 1}^m \sum_{j = 1}^n
\phi(f,\bm{x}^{+}_i,\bm{x}^{-}_j).
\end{equation}

Formally, we assume that $\bm{x}_0 \succ \bm{x}_1$ means that $\bm{x}_0$ should
be ranked above $\bm{x}_1$ while $\bm{x}_0 \prec \bm{x}_1$ means the opposite;
$\bm{x}_0 \equiv \bm{x}_1$ indicates they have the same importance. In the
multipartite ranking, we have the relation $\bm{x}_0 \succ \bm{x}_1$, $\bm{x}_0
\prec \bm{x}_1$ or $\bm{x}_0 \equiv \bm{x}_1$ for all pairs
$(\bm{x}_0,\bm{x}_1)$. The dataset $S$ can be divided into $L$ subset
$\{S_i\}_{i = 1}^L$, where $S = \cup S_i$ and $S_i \cap S_j = \phi$ for $i \neq
j$.

For $L$-partite ranking, the dataset $S$ can be divided into $\{S_i\}_{i =
1}^L$ according to the ratings of the instances. Generally, we can define the
empirical error or C-index \cite{Furnkranz2009} by extending
(\ref{eqn:bipartite_error}) to the multi-partite case:
\begin{equation}\label{eqn:multipartite_error}
R_{\phi}(f) =  \frac{1}{Z} \sum_{1 \le a < b \le L} \sum_{i = 1}^{|S_a|}
\sum_{j = 1}^{|S_b|}\phi(f,\bm{x}^{b}_i,\bm{x}^{a}_j),
\end{equation}
where $Z = \sum_{1 \le a < b \le L} |S_a||S_b|$. In the case of the
multipartite problem,  (\ref{eqn:r_f}) will be extended as C-index measure:
\begin{equation}\label{eqn:c_index}
R_{I}(f) =  \frac{1}{Z} \sum_{1 \le a < b \le L} \sum_{i = 1}^{|S_a|} \sum_{j =
1}^{|S_b|}  I[f(\bm{x}^{b}_i) < f(\bm{x}^{a}_j)],
\end{equation}

\subsection{Evaluation Measure}
In DC2010, evaluation is in terms of the NDCG (Normalized
Discounted Cumulative Gain) with the following ratings and the
discount function given the sorted ranking sequence $g$ and the
ratings of the instances (note that here $r_i \in \{0,1,\dots, L -
1\}$):
\begin{equation}\label{eqn:dc_ndcg}
DCG_g = \sum_{i = 1}^{|S|}r_i(|S| - i)\textrm{ , }NDCG =
\frac{1}{DCG_{\pi}}DCG_g,
\end{equation}
where $DCG_{\pi}$ is the normalization factor that is DCG in the
ideal permutation $\pi$ ($DCG_g \le DCG_{\pi}$).

\subsection{RankBoost for Ranking}
\begin{algorithm}
 \caption{RankBoost Algorithm}\label{alg:rankboostb}
\begin{algorithmic}
\STATE Input $(S^{b},S^{a}) \in X \times X$. \STATE Initialize
$D_1(\mathbf{x}^{b}_i,\mathbf{x}^{a}_j)$ for all $i \in \{1,2,\cdots,|S_b|\},j
\in \{1,2\cdots,|S_a|\}$. \FOR {$t = 1,2,\cdots,T$} \STATE  Train the weak
learner using the distribution $D_t$ to get weak ranker $h_t$ \STATE Choose
$\alpha_t \in R$ \STATE Update:
\begin{equation}
D_{t+1}(\mathbf{x}^{b}_i,\mathbf{x}^{a}_j) =
\frac{1}{Z_t}D_t(\mathbf{x}^{b}_i,\mathbf{x}^{a}_j)\exp(-\alpha_t(h_t(\mathbf{x}^{b}_i)
- h_t(\mathbf{x}^{a}_j))),
\end{equation}
where $Z_t = \sum_{i = 1}^{|S_b|} \sum_{j = 1}^{|S_a|}
D_t(\mathbf{x}^{b}_i,\mathbf{x}^{a}_j)\exp(-\alpha_t(h_t(\mathbf{x}^{b}_i) -
h_t(\mathbf{x}^{a}_j)))$
 \ENDFOR
\end{algorithmic}
Output the final ranking: $f(\mathbf{x}) = \sum_{t = 1}^T \alpha_t
h_t(\mathbf{x})$
\end{algorithm}

Rankboost \cite{Freund2003} maintains a distribution $D_t$ over $X \times X$
that is passed to the weak learner and approximate the true ordering by
combining many simple weak ranker. It minimize the following expected loss:
\eqn{}{ R_{rb}(f) =  \frac{1}{Z} \sum_{1 \le a < b \le L} \sum_{i = 1}^{|S_b|}
\sum_{j = 1}^{|S_a|}\exp(-[f(\mathbf{x}^{b}_i) - f(\mathbf{x}^{a}_j)]),} where
$f(\mathbf{x}) = \sum_{t = 1}^T h_t(\mathbf{x})$ and $1/Z$ may be regarded as
the probability on $(\mathbf{x}^{a}_i,\mathbf{x}^{b}_j)$ for all $i \in
\{1,2,\cdots,|S_a|\},j \in \{1,2\cdots,|S_b|\}$ in the initial distribution.

The weak ranker focuses on the binary rating ($0$ and $1$, or the positive and
the negative) that gives the relative ordering of the examples and ignores the
specific scores. The ranker has the following simple form:
\begin{equation}
h(x) = \left\{
  \begin{array}{ll}
    1, & \hbox{if $x_j \ge \theta$;} \\
    0, & \hbox{if $x_j < \theta$;} \\
    r_0, & \hbox{if $x_j$ missing,}
  \end{array}
\right.
\end{equation}
where $x_j$ is the j-th feature value of $\mathbf{x}$, $\theta$ is the
threshold and $r_0$ is the default value of the ranker. With the convexity of
$e^{\alpha x}$, it is easily verified that $((1 - r) e^{\alpha} + (1 +
r)e^{-\alpha})/2$ is the upper bound of $Z$ where
\begin{equation}
r = \sum_{\mathbf{x}^{b}_i \in S^{b},\mathbf{x}^{a}_j \in S^{a}}
D(\mathbf{x}^{b}_i,\mathbf{x}^{a}_j) (h(\mathbf{x}^{b}_i) -
h(\mathbf{x}^{a}_j)).
\end{equation}
The upper bound of $Z$ in each iteration is minimized when $\alpha =
\frac{1}{2} \ln ((1 + r)/(1 - r))$, which will yield $Z \le \sqrt{1 - r^2}$.
The weak ranker should choose the optimal value $j$, $\theta$ and $r_0$ to
maximize the weighted risk loss $|r|$. Algorithm \ref{alg:rankboostb} shows the
framework of RankBoost\footnote{For the bipartite ranking, the paper
\cite{Freund2003} gives a more efficient implementation called RankBoost.B}.

\section{Multi-Ranking with Encoding and Decoding}
In this section, we will decompose the multi-partite ranking into multiple bipartite ranking with the encoding and the decoding. In the coding stage, the ratings will be
coded as $k$-bit 0-1 sequence and each binary ranking algorithm may be the
point-wise ranking or the pair-wise ranking algorithm. In the decoding
stage, the final ranking score will be a weighted average of all binary ranking
algorithms and the descending sorted examples give the right rankings.

\subsection{Coding Design}
In this section, we describe the coding design for the rank learning. Given a
set of $L$ ratings to be learned and $L$ partite parts are formed. A codeword
of $n$ is designed for each rating, where each bit code represents the response
of a given dichotomizer \footnote{The dichotomizer generally denotes the binary
classifier in the literature of classification. Informally, here we introduce
this term into machine-learned ranking and call the bipartite ranker as the
dichotomizer.}. The codeword of each row is arranged to construct a coding
matrix $M_{L\times k}$, where $M_{ij} \in \{-1,0,1\}$. The l-th row
corresponding to the rating $l$ ($l = 0,1,\cdots,L - 1$) and the j-th column
the dichotomizer $h_j$ ($j = 1,2,\cdots,k$). We categorize the coding design
into the binary coding and the ternary coding basing on the range of the coding
value.

\subsubsection{Binary Coding}
\begin{figure}
  \centering
\includegraphics[width= 0.15\textwidth]{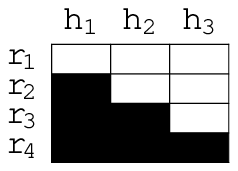}
   \caption{Binary coding design for 4 ratings (white: 0, black: 1)}
   \label{fig:binary_coding}
\end{figure}

The standard binary coding design is used for one-vs-all strategy in the
multi-class classification, where each dichotomizer is built to distinguish one
class from the rest of the classes. For the $L$-partite ranking problem, we
extend each rating to a vector with $L - 1$ dimensions. Formally, the coding
method will encode $l$ to $\bm{u}_l$ ($l = 0,1,\cdots,L - 1;j = 1,2,\cdots,L -
1$):
\begin{equation}\label{eqn:binary_coding}
u_{lj} = I[j \le l] = \left\{
        \begin{array}{ll}
          0, & \hbox{$j > l$;} \\
          1, & \hbox{$j \le l$.}
        \end{array}
      \right.
\end{equation}
Fig. \ref{fig:binary_coding} give the binary coding design with the
dichotomizer $\{h_1,h_2,h_3\}$ for 4 ratings. It can be explained with the fact
that the algorithm will sequently execute the dichotomizer $h_1,h_2,h_3$.
First, $h_1$ judge that it is hold that $r > 0$ for the instance $(x,r)$. Then,
$h_2,h_3$ test whether $r > 1$ and $r > 2$ or not, respectively. F\&H method
\cite{Frank2001} uses this encoding strategy to implement the ordinal
regression. Unlike their method, we use the pair-wise method (RankBoost) instead of the point-wise method. In practice, the pair-wise methods often achieve better performance than the point-wise methods.

\subsubsection{Ternary Coding}
\begin{figure}[t]
\centering
  \subfigure{
    \begin{minipage}[b]{0.15\textwidth}
      \centering
      \includegraphics[width= \textwidth]{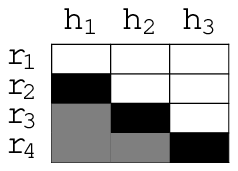}
    \end{minipage}}%
    \hspace{0.4in}
  \subfigure{
    \begin{minipage}[b]{0.15\textwidth}
      \centering
       \includegraphics[width= \textwidth]{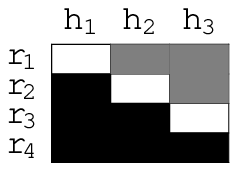}
    \end{minipage}}
  \caption{Ternary coding design for 4 rank values (white: 0, black: 1, gray: -1);
  left: $k$-column codes the instances from $\{S_i\}_{i = 1}^{k + 1}$;
  right: $k$-column codes the instance from $\{S_i\}_{i = k}^{L}$.}
  \label{fig:ternary_coding} 
\end{figure}

In classification, the ternary coding designs are the one-vs-one strategy
\cite{Hastie1998} and the sparse random strategy \cite{Allwein2000}. In
one-vs-one coding, the examples from all pairs of the classes are constructed
to train a model.  The negative code ($-1$) in the coding means a particular
class is not considered as the given classifier. We consider a $L - 1$-bit
coding design including the preference pairs, where the k-th column contains
the instances from $\{S_i\}_{i = 1}^{k + 1}$. Formally, we define the following
ternary coding ($l = 0,1,\cdots,L - 1;j = 1,2,\cdots,L - 1$):
\begin{equation}\label{eqn:ternary_coding}
u_{lj} = \left\{
           \begin{array}{ll}
             0, & \hbox{$l < j$;} \\
             1, & \hbox{$l = j$;} \\
             -1, & \hbox{$l > j$,}
           \end{array}
         \right.
\end{equation}
where the upper triangular part of the coding matrix is zeros called
the upper triangular coding. Optionally, the coding could also
contain all the instances from $\{S_i\}_{i = k}^{L}$ where the lower
triangular part of the coding matrix are 1 called the lower
triangular coding. Both of the ternary codings are depicted in Fig.
\ref{fig:ternary_coding}.

In Fig. \ref{fig:ternary_coding}, the matrix is coded as three dichotomizers
$\{h_1,h_2,h_3\}$ for the 4-partite problem. The white regions are coded by
$0$, the black regions by $1$, and the gray regions correspond to the $-1$ (the
regions that is not considered by the respective dichotomizer $h_j$). In the
left part of Fig. \ref{fig:ternary_coding}, the set of the partial relation
$\{(\bm{x}_0,\bm{x}_1)|\bm{x}_1 \in S_{j+1},\bm{x}_0 \prec \bm{x}_1\}$ will be
considered by the dichotomizer $h_j$.

\begin{figure}
  \centering
\includegraphics[width= 0.25\textwidth]{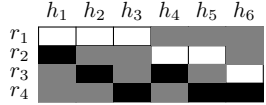}
   \caption{Ternary coding design for 4 ratings (white: 0, black: 1, gray: -1) in LPC and the model $f_{k,j}$
   corresponding to the pair of ratings $(k,j), 0 \le k < j \le L - 1$.}
   \label{fig:lpc_coding}
\end{figure}

We notice that LPC (Learning by Pairwise Comparison) \cite{Furnkranz2009} can be
formulated in the ternary coding with $L(L - 1)/2$ bits. As a example,  Fig.
\ref{fig:lpc_coding} show a coding with 6 bits when LPC solves a 4-partite
problem. The column $h_j$ corresponds to a pair of ratings $(l,k)$ such as $0
\le l < k \le L - 1$, where it only uses the examples with both of $l$ and $k$
ratings.

\subsection{Decoding Design}
In classification, the most frequently applied decoding designs are
built on the certain distance metric between the output vector and
the coding vector such as Hamming decoding \cite{Nilsson1965} for
the binary decoding and the loss-based decoding \cite{Allwein2000}
for the ternary decoding. ECOC will classify the instance to the
label whose coding vector is nearest to the output vector in the
specific distance metric. But for ranking, the objective of the
decoding design is to fuse the outputs of multiple dichotomizers
into a final ranking score instead of predicting the class label.
The procedure of the decoding design is to determine the weight of
each dichotomizer.

Recall that we assume the training dataset $\{(\bm{x}_i,r_i)\}_{i = 1}^{|S|}$,
where $r_i \in \{0,1,\cdots, L - 1\}$. Inspired by McRank \cite{Li2007} where
the ranking algorithm fused the posterior probability of the instance
conditioned on the class label into a score value, we define the scoring
function:
\begin{equation}
H(\bm{x}) = \sum_{k = 1}^{L - 1}T_k f(\bm{x}),
\end{equation}
where $T_k$ represents the weight function of the dichotomizer $h_k$. The
instances will be sorted in the descending order of $H(x)$ after computing the
weight scores for all instances. If we set $T_k = 1$ for all dichotomizers,
this is just as similar as the fusion manner of F\&H method. In our
experiments, we set both of $T_k = k - 1$ and $T_k = 1$ to implement our
algorithms. We find that it is obviously better to set $T_k = k - 1$ than $T_k
= 1$ if we assign the predefined weights to $T_k$. The linear transformation of
the scoring function will not change the ranking results. It seems that the
adaptive weighting function which measures the ability of the dichotomizers is
more intuitional than the predefined weighting. For each dichotomizer, we take
the $NDCG$ of 3-holdout validation instead of 3-crossfold validation as $T(k)$
considering the training time. In F\&H method, $T_{k}$ is set to 1 empirically.
LPC trains a seperate model $f_{k,j}$ for each pair of the ratings $(k,j)$ such
as $0 \le k < j \le L - 1$ and the prior probability $p_{k}p_{j}$ of the
ratings pair are used as the weights for the fusions: \eqn{}{H(\bm{x}) =
\sum_{0 \le k < j \le L - 1} p_{k} p_{j}f_{k,j}(\bm{x}).}

\section{Experiments}
\subsection{Description of Dataset}
In our experiments, we used all labeled samples of the English
language as training for the English quality task. DC2010 only
provides few labeled samples for the French and German tasks to
emphasize cross-lingual methods. We put all the labeled examples
including English, French and German into the training set for the
multilingual quality tasks (French and German tasks). We noticed
that due to the redirect of website there exist some duplicate
instances with the different ratings in the collection. In this
case, we chose the high ratings as the ranking value of the instance
and keep the unique instance in the dataset. After removing the
duplicated samples, we obtain the English training set with 2113
samples, French training set with 2334 samples, and German training
set with 2238 samples \footnote{We repeated to exact the features of
the datasets and found there was a little difference from the
results of our last competition (there exists some errors). But the
new results is comparable to the best results in the competition}.
The rating of the instance ranges from $0$ to $9$, which is measured
as an aggregate function of genre, trust, factuality and bias and
spam and is more delicate than the LETOR dataset \cite{Qin2010}
(with 3 ratings).

The test dataset includes $31,893$ English instances, $5,337$ French
instances and $19,564$ test instances. The organizations of the
competition provided the instances with their ratings sampled from
the test dataset to the participants for optimizing the ranking
algorithms. Then they would extract another group of instances
randomly from the test dataset to test the ranking algorithms as the
final competition results. Tab. \ref{tab:dataset_symbol} gives the
descriptions of the sampling set from the test dataset, where six
sampling set are denoted as $D_i,i=1,2,3,4,5,6$
\begin{table}[!htp]
  \centering
  \caption{Description of dataset}\label{tab:dataset_symbol}
  \begin{tabular}{|l|c|c|c|}
    \hline
    Dataset & Symbol & training & test \\
    \hline
    English sampling& $D_1$ & 2,113 & 131 \\
    \hline
    English final& $D_2$ & 2,113 & 1,314  \\
    \hline
    Germany sampling& $D_3$ & 2,334 &75  \\
    \hline
    Germany final& $D_4$ & 2,334 & 234 \\
    \hline
    French sampling& $D_5$ & 2,238 & 138 \\
    \hline
    French final& $D_6$ &2,238 &274  \\
    \hline
  \end{tabular}
\end{table}

\subsection{Experiment Results and Discussions}
\begin{figure}[!htp]
\centering
\includegraphics[angle=0, width=0.4\textwidth]{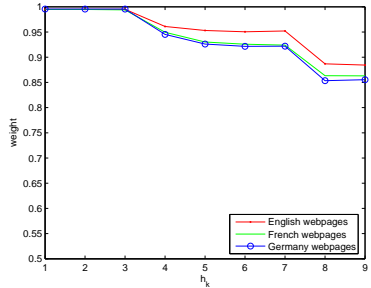}
\caption{During the decoding with the adaptive weighting, weights
change with the dichotomizer $h_k$ for the ranking tasks of three
languages}\label{fig:ada_weight}
\end{figure}

In the following experiments, we set the number of the weak ranker
to 100 and use all attributes as the thresholds of the weak ranker
if not special specified.

For the bipartite ranking problem, the ranker should rank the
positive instances in the head of the rank sequence as far as
possible. The rank problem will become easier and easier when the
ratio of the positive instances relative to the negative ones
increases. In the extreme case, any permutation will be regarded as
correct while all instances are positive. Fig. \ref{fig:ada_weight}
shows the holdout NDCG (as the adaptive weighting) for each
dichotomizer under the binary coding. For the binary coding, the
ratios of the instances decreases with the increasing $k$ and the
performance of the dichotomizer decrease step by step with $k$.

\begin{figure}[!htp]
\centering
\includegraphics[angle=0, width=0.4\textwidth]{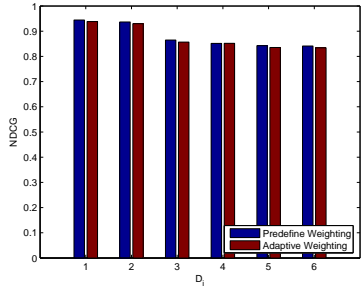}
\caption{Comparisons among two decoding method with the binary
coding}\label{fig:ndcg_bin_pre_ada}
\end{figure}
How can we improve the overall performance according to the real
situation of the dichotomizer? The critical view is to compensate
the dichotomizer with the large $k$. In the 4-partite ranking
problem, a instance with ratings 3 in the dichotomizer $h_3$
probably can not be ranked in the head of the sequence due to its
encountering disadvantage situation. But in other dichotomizers
which have more perfect performance than $h_3$, a instance with
ratings 3 likely is ranked ahead. The instance with ratings 3 will
keep its advantage when giving it a high weight for the
compensation. Figs. \ref{fig:ndcg_bin_pre_ada} give a comparison
between the predefine weighting and the adaptive weighting under the
binary coding manner, where the decreasing weighting means a
negative compensation and gives the inferior performance compared
with the predefine weighting.

\begin{figure*}[!htp]
\centering
  \subfigure{
    \begin{minipage}[b]{0.4\textwidth}
      \centering
\includegraphics[angle=0,width= \textwidth]{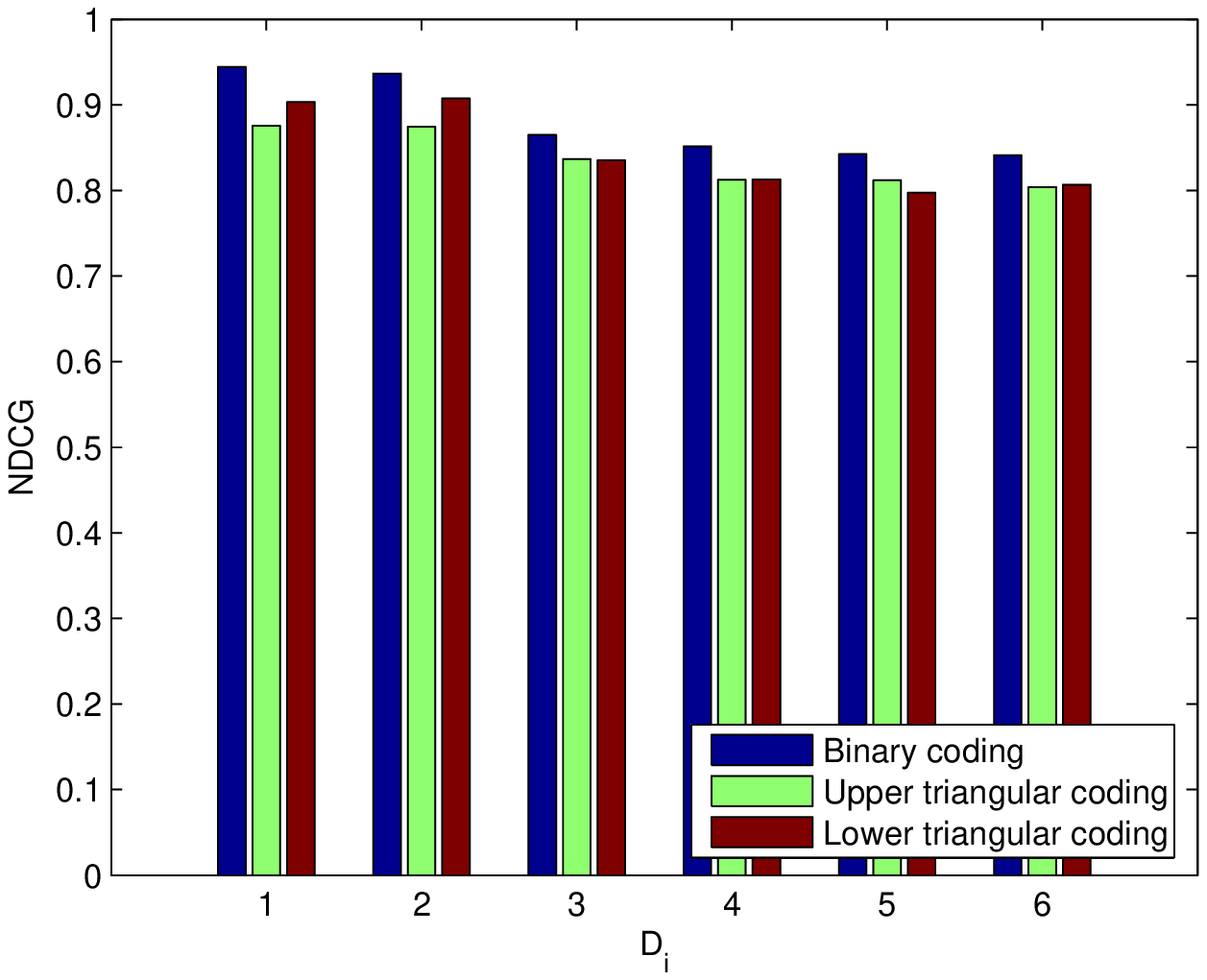}
    \end{minipage}}%
  \subfigure{
    \begin{minipage}[b]{0.4\textwidth}
      \centering
       \includegraphics[angle=0,width= \textwidth]{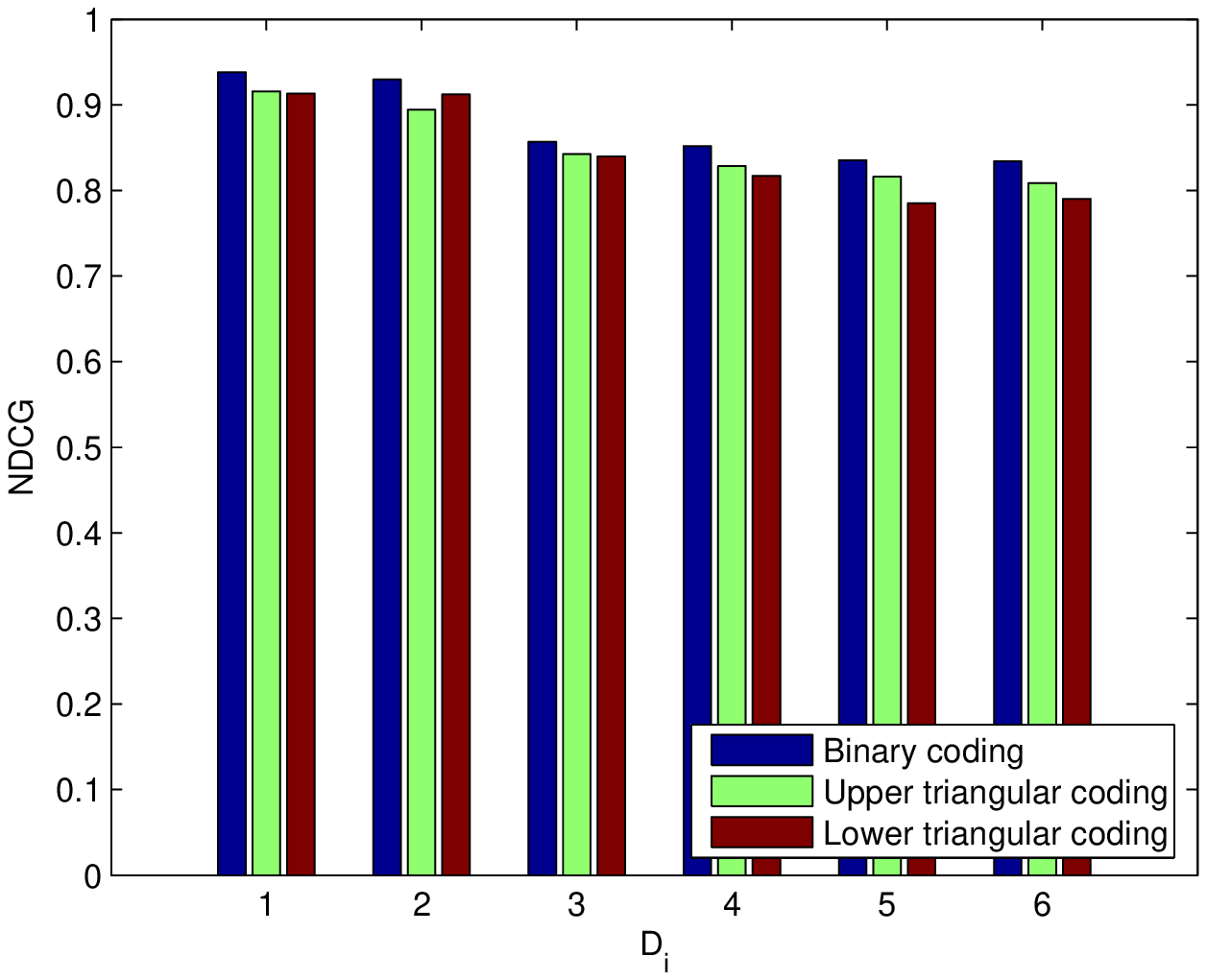}
    \end{minipage}}
  \caption{Comparisons among three coding methods with the two decoding. Left: the predefine weighting,
  right: the adaptive weighting}
  \label{fig:ndcg_bin_tri_ada_pre} 
\end{figure*}

Fig. \ref{fig:ndcg_bin_tri_ada_pre} give the comparisons among three
coding methods under two decodings method. In both of the predefine
weighting and the adaptive weighting, we can see that the binary
coding outperforms the ternary coding. Moreover, let
us compare the lower triangular coding and the upper triangular
coding.  It is interesting that the lower triangular coding is more
effective than the upper triangular coding for the predefine
weighting decoding and the opposite case holds for the adaptive
weighting decoding. Compared with the upper triangular coding, the
lower triangular coding is prone to be disturbed by the weight
compensation.

\begin{figure}[!htp]
\centering
\includegraphics[angle=0, width=0.4\textwidth]{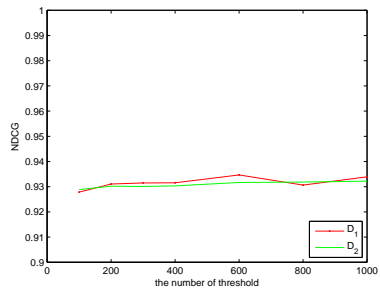}
\caption{The NDCG measurer changes with the number of the
thresholds}\label{fig:threshold_ndcg}
\end{figure}

The weak ranker determines output $0$ or $1$ by comparing the
specified attribute of the instance with the threshold. A candidate
set of the thresholds can be provided for searching the optimal
threshold on the dataset. Fig. \ref{fig:threshold_ndcg} show that
the NDCG measure increases slightly when given more thresholds,
where the horizontal axis represents the number of the thresholds
for each attribute. It can be explained with the fact that the
algorithm will obtain a more optimal threshold when enlarging the
searching region.

\begin{figure}[!htp]
\centering
\includegraphics[angle=0, width=0.4\textwidth]{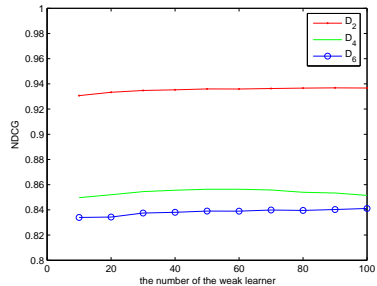}
\caption{The performance of MultiRanking.ED with the binary coding and the
predefine weighting decoding change with the number of the
 weak learner.}\label{fig:ecoc_number_weak_learner}
\end{figure}

In classification, AdaBoost \cite{Reyzin2006} methods are known not
to usually overfit training data even as the size of the weak
classifiers becomes large. RankBoost can be regarded as the
application of the AdaBoost in the pairs of the instances. In Fig.
\ref{fig:ecoc_number_weak_learner}, we see that the NDCGs vary
gently (behaving nearly identically) and resist overfitting, which
is consistent with \cite{Freund2003}.

\begin{table}[htbp]
\caption{Bagging+C4.5 and MultiRanking.ED are compared on six
datasets. MultiRanking.ED adopts the binary coding and the predefine weighting
decoding. The numbers in the brace of the columns represent
the number of the thresholds. The second column represents using all threshold for the weak learners. For all results, the number of the weak
learner is set to 100.}\label{tab:ecocrankboost}
\begin{tabular}{|l|c|c|c|c|}
\hline
  dataset &Bagging+  &MultiRanking  &  MultiRanking  & MultiRanking \\
        & C4.5 & .ED &.ED(100) &.ED(1000)  \\
  \hline
$D_1$&0.9325&$\bm{0.9442}$&0.9279&0.9339 \\
\hline
$D_2$&$\bm{0.9378}$&0.9367&0.9289&0.9322\\
\hline
$D_3$&0.8620&$0.8649$&0.8610&$\bm{0.8657}$\\
\hline
$D_4$&0.8484&$\bm{0.8515}$&0.8515&0.8502\\
\hline
$D_5$&0.8359&$0.8425$&$\bm{0.8430}$&0.8397\\
\hline
$D_6$&0.8405&0.8411&$\bm{0.8424}$&0.8402\\
\hline
\end{tabular}
\end{table}

Finally, Tab. \ref{tab:ecocrankboost} gives the comparisons between
Bagging + C4.5 and MultiRanking.ED, which adopts
the binary coding and the predefine weighting decoding. Comparing
the first column and the second column, all rows except the second
row show that MultiRanking.ED gives a better performance than Bagging +
C4.5, which is the best competition results in DC2010.The middle two columns also present the performance of MultiRanking.ED under the different number of the
thresholds.

\section{Conclusion}
In this study, we try to solve the web quality evaluation problems with the combination of multiple bipartite pair-wise ranking models by the efficient encoding and decoding strategy.In coding, we present the
binary coding and the ternary coding extending the ratings to $L -
1$ dimension vector. For decoding, we give the combination of the
ranking sequences with the predefine weighting and the adaptive
weighting. The ECML/PKDD Discovery Challenge
2010 datasets provide a chance to validate our proposed algorithm,
which contains English, French and Germany webpage quality tasks. We dicussed the probable factors which influence the NDCG measurer including the number of weakers, the number of the thresholds and the different encoding and the decoding strategy experimentally.
The final results show that our algorithm MultiRanking.ED with the binary coding and the predefine weighting decoding  overpasses its counterparts and gives a perfect performance.

In future work, we will explore the fashions to combine the multiple
rank sequence effectively.  Another direction is to validate it on other state-of-art datasets.

\section*{Acknowledgments}
This work is supported in part by Innovation Scientists and
Technicians Troop Construction Projects of Henan Province
under grants No.094200510009 and the National Natural Science Foundation of China (NSFC) under Grant No. 61103138 and No. 61005029.

\bibliographystyle{elsarticle-num}

\end{document}